\documentclass[12pt,aps,prb,preprint]{revtex4}   

\begin{document}

\title{Difference Principle and Black-hole Thermodynamics}

\author{Pete Martin}

\affiliation{Spatial Information Research Centre, Department of
Information Science, University of Otago, Dunedin, New Zealand}

 \email{pmartin@ieee.org}

\date{\today}

\begin{abstract}
The heuristic principle that constructive dynamics may arise
wherever there exists a difference, or gradient, is discussed.
Consideration of black-hole entropy appears to provide a clue for
setting a lower bound on any extensive measure of such collective
system difference, or potential to give rise to constructive
dynamics. It is seen that the second-power dependence of
black-hole entropy on mass is consistent with the difference
principle, while consideration of Hawking radiation forces one to
beware of implicit figure-ground distinctions in the application
of the difference principle.\end{abstract}


\maketitle

\section{Introduction}
In this paper a general physical principle, called the
``difference principle'', is presented. This principle may be used
to gain insight into questions of dynamics in many contexts, for
purposes of instruction or the conception of theory. The
difference principle is an expression of the observation that
difference in space (non-equilibrium) bears a reciprocal and
self-propagating relation to difference in time (flow, or change).

The structure of the paper is as follows: Section
\ref{Metaphysics} is a motivation and justification for
presentation of ``principles'' in general (as contrasted to
specific scientific theories).  Section \ref{Examples} comprises
the main exposition via examples of the principle in its relation to
the general field of non-equilibrium thermodynamics. Section
\ref{BlackHole} is a discussion of the difference principle in
relation to black-hole thermodynamics.

\section{Physical Laws, Principles, and Logic}\label{Metaphysics}

Our interest in physics begins with innocent, reasonable questions
like: How can a sailboat go upwind? How can an albatross keep
flying, without ever beating its wings? How can a burning flame
make the inside of a gas refrigerator cold? These are presumably
the more advanced questions that follow the more fundamental
questions of the type: What makes the wind blow? Why does heat go
from hot places to cold places?

What's in an explanation? In his formal theory of inductive
inference, Solomonoff proposed that scientific theory seeks
information compression, and that scientific laws essentially
summarize observations (including observations to come) in compact
form.\cite{Solomonoff:1964} With this view, the essence of science
is to discover the correlations of the world; there is no mandate
to provide account of a temporal causal sequence that comprises
the history of an observation in need of explanation. Perhaps this
is in consonance with the ``Ithaca interpretation'' of quantum
mechanics that  ``correlations have physical reality; that which
they correlate does not'',\cite{Mermin:1998} or Wheeler's ``it
from bit'' program to discover the possible information-theoretic
foundation of all physics.\cite{Wheeler:1990}

Classical physics, the physics of our experience and presumably
the physics that influenced the evolution of our cognition, is
evidently local, causal, and reversible at the microscopic
level.\cite{Feynman:1982} Perhaps for this reason we favor causal
accounts of observations, even though our everyday experience
depends in a very essential way on statistical
effects,\cite{Schrodinger:statistics} and even though the leading
edge of theoretical physics forces us to consider that there may
be no fundamental causal laws, but only probabilities for physical
processes,\cite{Anandan:2003} or that the ultimate causal laws may
not be simple at all, so that the best we can do is to summarize
the ``emergent'' patterns and rhythms that we observe, effectively
assuming that the underlying microscopic dynamics, at the finest
scale, are ``random''  for all practical
purposes.\cite{Nielsen:1987}

Our preference for causal, local explanation is probably the
reason for the predominance of Newton's vectorial mechanics (based
on the vector quantities \emph{momentum} and \emph{force}) over
the mathematically equivalent analytic mechanics of Leibniz,
Euler, and Lagrange (based on the scalar quantities \emph{energy}
and \emph{action} (see, for example, Lanczos\cite{Lanczos:1962}), to
which Hamilton's action principle belongs.

When Feynman presented his alternative derivation of quantum
mechanics based on path integrals, inspired by thinking about the
action principle, he disavowed having presented any fundamentally
new results, but noted that ``there is a pleasure in recognizing
old things from a new point of view''.\cite{Feynman:1948}  There
is also great heuristic value in this, insofar as familiarity with
various points of view may lead to fundamentally new results.

Even if certain principles of physics are not considered to be
laws or explanations in their own right, they may serve the
purpose of guiding one to correct hypotheses. The principle that a
beam of light should take the quickest (not the shortest) path
between two given points, for example, does not provide a local,
causal account of how the light is refracted or ``knows'' what
path to take, but it leads to a correct formulation of geometric
optics. The Le Chatelier-Braun principle, that a chemical balance
responds to an externally-imposed change of temperature, pressure,
etc., in such a way as to counteract the imposed change, offers no
mechanistic account of the process, but qualitatively summarizes a
class of generally observed behavior, leading to proper
expectations regarding further observations, and to good guesses
for the construction of specific hypotheses.

The principle of evolution by natural selection exemplifies the
quasi-tautological nature of certain scientific principles. The
biologist Lotka observed in 1922 that the principle of natural
selection might be considered to be a physical
principle.\cite{Lotka:1922b} When stated as ``one should expect to
observe that which is able to propagate into the future,'' the
principle appears to be almost pure logic.

Notably, the second law of thermodynamics, which is closely
related to the difference principle that I discuss here, is
regarded by many not to be a physical law at all, but rather to
belong to the realm of logical scientific
inference.\cite{Cox:1946, Jaynes:1957, Jaynes:1965, Toffoli:2003}
Penrose has addressed the question of time asymmetry between
probabilistic inference applied to the past versus probabilistic
inference applied to the future, and shown the second law of
thermodynamics to be a consequence of an extraordinarily
low-entropy state of the early universe. In this paper the
assumption is made that there is a subtle but useful distinction
to be made between epistemological content, on the one hand, and
the ontological content of what Toffoli has called
``substrate-universal behavior'',\cite{Toffoli:2003} on the other.
Whereas the ``entropic dynamics'' program of
Caticha\cite{Caticha:2001, Caticha:2008} aims to derive the laws
of dynamics purely from rules of inference, here I use the phrase
``entropy dynamics'' to refer to the fundamental
difference-reduction and difference-generation reciprocal
processes of Nature that, after all, gave rise to our capacity for
mental models and to our particular form of logic.

Analogy is the core of cognition, according to
Hofstadter.\cite{Hofstadter:2002} In learning and modeling our
world, we construct larger and larger equivalence classes of
phenomena in a process of conceptual unification that he calls
``chunking''. Following this view one could say that, just as
there is pleasure in seeing old things in a new way, conversely
there is pleasure in recognizing how new things belong to enlarged
revisions of old equivalence classes.

Whether or not we can say exactly why, we soon grow accustomed to
the equivalence class of phenomena comprising things that go
``downhill'', whether by freedom (diffusion) or by law (field).
Later we might become familiar with the equivalence class of
phenomena comprising things that go uphill whenever something else
goes downhill. Such is the class of phenomena to which the opening
questions refer. In constructing this equivalence class, we learn
to appreciate that the discharge of one difference may be
leveraged to generate another difference, and that in fact this
reciprocal relation may be a prevailing theme in the natural
world, as much as in the realm of human ingenuity.

\section{The Difference Principle in Entropy Dynamics}\label{Examples}

\begin{quote}
Where a difference of temperature exists, motive force can be
produced.
 \end{quote}

This was the  observation of Sadi Carnot,\cite{Carnot:1977} one of
the founders of thermodynamics, before any formal statement of the
Second Law of thermodynamics, and before the introduction of the
term ``entropy'' by Clausius.  One might paraphrase Carnot's
principle by saying that the flow of heat ``downhill'' from a
high-temperature source to a low-temperature sink can be leveraged
to generate a mechanical potential.

Of course it is analogously true that where a difference of
elevation exists (as of hydraulic head) motive force can be
produced, or that where a difference of chemical potential exists,
motive force can be produced.

In seeking to explain the \emph{why} of dynamics, as well as of
various sorts of organization or ``order'' apparent in our
environment, it is customary to give an energy account.  We might,
for example, trace a certain chemical potential  upon which our
society depends back to the constructive activities of organisms
dependent upon the free energy of the sun, which in turn derives
from the free energy of gravitational potential. ``The flow of
energy through a system tends to organize that
system''\cite{Morowitz:1968} is a well-known general principle.

One might observe the organizing effect of tidal flow in an
estuary, and take account of its ultimate origin in angular
momentum. But apparently the effect depends upon the
\emph{equilibration} of the \emph{difference} in angular velocity
between the Earth and the Moon; angular momentum per se is of less
importance than the difference. Perhaps we would be justified to
revise the above-mentioned energy-flow principle:  ``The
equilibration of a difference through a system tends to organize
that system''. It is noteworthy that, from a structural point of
view rather than from a functional point of view,
\emph{organization} is \emph{differentiation} (including the
creation of hierarchy).

\subsection{Difference and Motive}

\begin{quote}
There can be no distinction without motive, and there can be no
motive unless contents are seen to differ in value.
\end{quote}

So began George Spencer-Brown's development of symbolic logic from
more primitive concepts.\cite{Spencer-Brown:motive} His work lay
outside of physics as usually defined, yet the statement applies
to physical processes, as a description of substrate-universal
behavior.

We recognize that the latent motive inherent in the hydraulic head
of a river derives from the energy of the sun, and we observe that
the resulting flow of the river may generate a succeeding
differentiation in the form of sorted alluvium in placer deposits.
The discharge of the water's hydraulic (gravitational) potential
might be used to generate electrical potential; current from the
electrical potential may drive a motor, which drives a pump, and
so forth in a cascade of sequentially-orthogonal differences and
flows.

Here ``sequentially-orthogonal'' is used to mean that the
succeeding difference generated by the discharge of the preceding
difference may appear in a state-space dimension orthogonal to
that of the preceding difference, as for example electrical
potential could be considered to be orthogonal to gravitational
potential, because they are mutually independent, though each can
be transformed into the other. Clearly the second law of
thermodynamics precludes reversible transformation back and forth
without net equilibration of difference, but the essential point
is that the second law exhibits a constructive aspect in the
operation of entropy dynamics---the complementary processes of
equilibration of differences and generation of orthogonal
differences.

``Asymmetry is the cause of phenomena,''  Pierre Curie
said.\cite{Curie:1894} Asymmetry is essentially difference.
Symmetry, Rosen shows us, is akin to entropy.\cite{Rosen:entropy}
Some reflection suggests that Carnot's principle could have been
stated with greater generality, as an expression of the difference
principle of entropy dynamics:

\begin{quote}
Where a difference exists, motive force can be produced.
\end{quote}
The principle is not, after all, essentially dependent upon the
specific nature of the primary difference.

It also happens that a difference of temperature can, by way of
heat flow, produce an electrical potential (the so-called Seebeck
effect), while a difference of electrical potential can, by way of
electrical current, produce a thermal potential (the so-called
Peltier effect); Kelvin conjectured, and Onsager confirmed, that
these reciprocal relations were quantitatively
symmetrical.\cite{Onsager:1931}

A more complete expression of the difference principle of entropy
dynamics might therefore be given as:
\begin{quote}
Where a difference exists, motive force can be produced, and where
a motive force exists, a difference can be produced.
\end{quote}
Spatial and temporal difference appear as self-propagating duals.
The use of the term ``dual'' is intended to echo Toffoli's use of
the word when he conjectured that \emph{entropy} is a concept dual
to that of \emph{action}:\cite{Toffoli:1998} as \emph{entropy} is
interpreted as ``uncertainty as to state'', so \emph{action} is
interpreted to be ``uncertainty as to trajectory''. Likewise I
conjecture that any measure of collective internal
\emph{difference} of state, that is, any measure of
non-equilibrium or asymmetry, should be associated with a dual
measure of \emph{change}, or that collective characteristic of
trajectories that would imply the subsequent generation of further
non-equilibrium.

\subsection{Application of the Difference Principle}

So how can a sailboat go upwind? The value of the difference
principle in framing the problem is that it draws one's attention
to the asymmetry inherent in a system comprising air that is
moving and a substrate that is not. It leads to the expectation
that this asymmetry could be leveraged to generate orthogonal
asymmetry, and in a sequence of at least two steps, to produce a
``flow'' opposite the original gradient.

While the standard vector analysis of resultant force on a sail
together with the reaction of the keel explains the mechanics of
the mystery of making way against the wind, it helps little toward
understanding that one could just as well build a little
windmill-driven car that could advance over the ground directly
upwind. By reference to the difference principle, this possibility
would be obvious. Of course the ``efficiency'' of such
transformations of asymmetry are necessarily less than 100\%, so
the little car could not be expected to advance by its own induced
headwind!

How can an albatross keep flying, without ever beating its wings?
Understanding the difference principle, one might realize that,
whereas there is no inherent ``difference'' in a constant wind
field, if there is a wind gradient, then ``motive force'' can be
produced.

Suppose that there is a horizontal gradient, with the wind
velocity increasing with altitude above the seas. Then the
albatross could descend in the downwind direction, and ascend in
the upwind direction (by its own inertia), encountering both ways
an increasing supply of wind, so to speak, to make up for its loss
of airspeed due to friction. Hence a surplus of lift might be
gained, and the bird's sum of potential and kinetic energy
maintained or increased, at the expense of the wind gradient
(which gradient should be expected to decrease slightly, reducing
the wind speed aloft compared to that below).

Alternatively one might imagine riding in a hot-air balloon in
such a horizontal wind gradient, with a little wind generator on
board. Your only control of the balloon is to ascend or descend.
If you maintain constant altitude, you feel no relative wind, and
your wind generator can extract no work from the environment. It
makes no difference to you for this purpose that you may be moving
over the ground. But whenever you ascend or descend, your wind
generator works: you encounter the difference in the wind field,
and a motive force can be produced.

How can a burning flame make the inside of a gas refrigerator
cold? Again, the way that the difference principle might help to
frame the problem of understanding absorption-cycle refrigeration
is to recognize that the process must, as always, involve an
original source-and-sink difference, and it must involve at least
two orthogonal differences, or asymmetries. In the light of
familiarity with the difference principle in relation to entropy
dynamics, there is no reason to doubt that a temperature
difference could produce another temperature difference, via
intermediate transformations of asymmetry. In this case the
orthogonal intermediate asymmetry is that of chemical species
separated by heating. When this difference is discharged, by
allowing the species to mix, this flow can pump heat, producing
the difference in temperature between the inside of the
refrigerator and the outside.

So-called ``transitional dynamics'' are exploited by fish in their
swimming, by bees in their wing aerodynamics, and by engineers
designing bio-inspired propellors (see, for example,
\cite{Triantafyllou:1995, TriantafyllouEtAl:2004, UsabEtAl:2004}):
fish seek vortices and other irregularities in water flow that
they may leverage to their advantage, while bees and
underwater-vehicle propulsion engineers achieve greater
aerodynamic- or hydrodynamic efficiency with foils operating
across the dynamic flow-regime boundary of stall. Even stochastic
resonance may be seen to fit the ``difference principle''
paradigm; as Berdichevsky and Gitterman observe, ``Common to
ratchets and stochastic resonance is their ability to extract
useful work from a fluctuating
environment''.\cite{BerdichevskyGitterman:1998}\footnote{Stochastic
resonance is the effect that a signal lying below the threshold of
detection may become detectable by the addition of noise.}

As a further illustration of the ``difference principle'' way of
looking at a problem, I offer the following example of a process
involving sequentially-orthogonal differences propagated via the
duals \emph{potential} and \emph{flow}. When an electrical
engineer has available a certain dc electrical potential and would
like to produce a higher dc electrical potential, one scheme that
would come to mind would be to arrange a circuit wherein a
changing current, which is of course discharge of electrical
potential, produces a changing magnetic potential (orthogonal to
the electrical potential). The changing magnetic potential in turn
produces the secondary electrical potential.  In simple practical
terms, the circuit is an oscillator and a transformer followed by
a rectifier (an electrical ratchet). The value of the difference
principle is simply that it guides one to expect that the solution
should exist, and that it may involve a two-step sequence of
orthogonal differences.

In contrast, a similar example that does not appear to involve
orthogonal differences is the case of the water ram. Here
relatively low hydraulic potential is discharged, through a pipe,
and the flow is periodically arrested, so that the inertial
``water hammer'' effect can be used, with a check valve (water
ratchet), to produce relatively high hydraulic potential. But the
secondary hydraulic potential can be considered to be orthogonal
to the primary hydraulic potential insofar as the two potentials
are independent (aside from their coupling through the mechanism
itself).

\section{Black-hole Thermodynamics}\label{BlackHole}

In the Penrose account of entropy dynamics, the original great
state of non-equilibrium of the early universe was its uniform
distribution of mass-energy (near-zero Weyl tensor of space-time
curvature), providing, in effect, maximal collective gravitational
potential.\cite{Penrose:thermodynamics}  In the interest of
finding a theoretical lower bound for collective internal system
``difference'', non-equilibrium, or potential for further
constructive dynamics, we try to imagine the opposite of uniform
distribution of mass-energy, which would be singular concentration
of mass-energy: we consider black holes,  following the lines of
thought of Bekenstein\cite{Bekenstein:1981} and
Lloyd\cite{Lloyd:2000}. The presumption is that a black hole can
be taken as the physical realization of a ``dead
battery''\footnote{What Americans call a ``dead'' battery, others
call a ``flat'' battery, quite appropriately, since ``flat''
describes a high-entropy distribution.}, compared to the
``fully-charged battery'' of the low-entropy early universe
configuration\footnote{Interestingly, \emph{flat} space has the
least entropy, presumably for combinatorial reasons.}.

\subsection{Second-power mass relation}

The Bekenstein-Hawking entropy of a black hole $S_{BH}$ is given
by
\begin{equation}\label{BlackHoleThermEnt}
S_{BH} = \frac{kAc^{3}}{4G\hbar},
\end{equation}
where $k$ is Boltzmann's constant, $A$ is the area of the event
horizon, $c$ is the speed of light, $G$ is the gravitation
constant, and $\hbar$ is Planck's constant divided by $2\pi$. As
shown by Bekenstein,\cite{Bekenstein:1973} the entropy of a black-hole is
proportional to its surface area.

 Since the Schwarzschild radius
$R_{S}$ is given by
\begin{equation}\label{SchwarzRad}
R_{S} = \frac{2GM_{BH}}{c^{2}},
\end{equation}
where $M_{BH}$ is the mass, combining eq. (\ref{BlackHoleThermEnt})
and eq. (\ref{SchwarzRad}), and taking $A = 4\pi R_{S}^{2}$, the
black-hole thermodynamic entropy can be expressed as a function of
mass:
\begin{equation}
S_{BH} = \frac{4\pi k G M_{BH}^{2}}{\hbar c}.
\end{equation}
Thus the thermodynamic entropy is proportional to the
\emph{square} of the mass (or energy).

Boltzmann's equation relates thermodynamic entropy $S$ to
information by
\begin{equation}\label{BoltzEnt}
S = k\ln(W),
\end{equation}
where $W$ is the number of microstates consistent with the
macrostate for which entropy is to be quantified; $\ln(W)$ is
therefore the information (uncertainty as to state) in ``nats'',
and dividing this by $\ln(2)$ would express the information in
bits. Hence combining eq. (\ref{BlackHoleThermEnt}) and eq. (\ref{BoltzEnt}), the information entropy of a black hole $I_{BH}$
in bits, as a function of the event horizon area $A$, is given
(see for example Wheeler\cite{Wheeler:1990}) by
\begin{equation}
I_{BH} = \frac{A c^{3}}{4\ln(2) G \hbar},
\end{equation}
or as a function of the mass $M_{BH}$  by
\begin{equation}
I _{BH} = \frac{4 \pi G M_{BH}^{2}}{\ln(2) \hbar c}.
\end{equation}

It seems a bit surprising that the upper bound of the (missing)
information content of a system would be proportional to the area
of its boundary, and to the \emph{square} of its mass, or energy,
which might otherwise be considered to be the natural choice for a
measure of system ``size''. After all, the upper bound for the
speed of dynamical evolution of a system has been shown by
Margolus and Levitin to be directly proportional to the average
energy of the system:\cite{MargolusLevitin:1998}
\begin{equation}
f \leq \frac{E}{\pi \hbar},
\end{equation}
where $f$ is the number of mutually orthogonal changes per second,
and $E$ is the average system energy.

Perhaps the most surprising consequence of the second-power
relation between mass and black-hole entropy, as noted by
Bekenstein, is that the entropy of a black hole formed by merging
two separate black holes is greater than the sum of the separate
black hole entropies, for the simple reason that $(M_{1} +
M_{2})^{2} \geq M_{1}^{2} + M_{2}^{2}$: the entropy of the single
black hole resulting from the merging of two similar, separate,
black holes is double the sum of the entropies of the two black
holes taken separately.

A physical consequence of this is that work could be done by
merging two black holes, perhaps by extracting energy from
gravitational waves.\cite{Hawking:1971} This seems as odd as if
one could wring further life out of two dead batteries by
connecting something between them. And it seems to fly in the face
of the ``difference principle'', or its converse, which would
state, after the Spencer-Brown statement  ``there can be no motive
unless contents are seen to differ in value'', that
\begin{quote}
Where no difference exists, no motive force can be produced.
\end{quote}

But in fact careful consideration of the difference principle
reveals that it would lead to just the correct expectation,
because it refers to any \emph{internal} difference inherent to
the system under consideration. Whereas a single black hole,
considered as a system (bounded perhaps by its event horizon),
apparently has exhausted its potential for further constructive
evolution, a pair of black holes must be considered as a larger
system including the very difference posited by the existence of
two distinguishable black holes.

The difference principle would lead one to expect that, in a
system comprising more than one object, or indeed in a space
comprising two distinguishable regions, there would be the
possibility for further change---a ``motive force'' could be
produced. If one tentatively hypothesized that black-hole entropy,
as a sort of negative proxy for system internal non-equilibrium,
were proportional to the first power of mass, and if one accepted
the conservation of mass, then one would find in the merging of
two black holes that the difference principle would be violated,
insofar as the prior existence of two distinguishable black holes
``made no difference'' when they merged to become one; entropy
remained constant. Guided then by the difference principle to
expect a non-linear functional relation $f$ between mass $M_{BH}$
and entropy $S_{BH}$ for which
\begin{equation}
f(M_{1} + M_{2}) \geq f(M_{1}) + f(M_{2}),
\end{equation}
one might be led to the simple guess that
\begin{equation}
S_{BH}\propto M_{BH}^{2}.
\end{equation}

\subsection{Hawking radiation}

A more serious challenge to the applicability of the difference
principle to black-hole thermodynamics is Hawking radiation.
Thermodynamic entropy has the units of energy divided by
temperature; if an object has energy and entropy, one might expect
that the object should also have a temperature. Hawking was thus
led to the discovery that a black hole does have a characteristic
temperature, and can be expected to emit black-body radiation
corresponding to this temperature.\cite{Hawking:1976}

The temperature $T_{BH}$ associated with a black hole is given by
Hawking:\cite{Hawking:1974}
\begin{equation}\label{HawkingTemp}
T_{BH} = \frac{\kappa \hbar}{4\pi k},
\end{equation}
where $\kappa$ is the surface gravity, which we will take to be
given by
\begin{equation}\label{SurfGrav}
\kappa = \frac{G M_{BH}}{R_{S}^{2}}, \end{equation} so that,
combining eqs. (\ref{SchwarzRad}), (\ref{HawkingTemp}), and
(\ref{SurfGrav}), black hole temperature may be expressed as a
function of mass:
\begin{equation}\label{TempMass}
T_{BH} = \frac{c^{4} \hbar}{16 \pi k G M_{BH}}.
\end{equation}
Therefore $T_{BH}$ is inversely proportional to $M_{BH}$, and the
black hole is seen to have negative specific
heat.\cite{Hawking:1976}

The black hole was presumed to be at maximum entropy for a system
of that mass---``exhausted'' so to speak, with respect to the
possibility for further constructive evolution---yet evidently
there was the potential for an encore: it could
explode!\cite{Hawking:1974} In the absence of accretion from its
surroundings, a black hole emits Hawking radiation at an
increasing rate, since it has negative specific heat and its
temperature increases as it sheds mass. Even in the absence of
accretion from its surroundings, it turns out that a black hole of
one solar mass could endure longer than the present age of the
universe (its theoretical lifetime being about $10^{64}$
years,\cite{Hawking:1974}) but a black hole of 1kg mass would blow
up in a matter of $10^{-21}$ second.\cite{LloydNg:2004} Certainly
such a 100-megaton explosion represents further dynamics, possibly
even constructive dynamics in the sense that this flux of energy,
this ``motive force'', is likely to produce a new difference
somewhere in the whole system considered.

``The whole system considered'' is the key phrase to the
resolution of the apparent paradox of Hawking radiation and
black-hole explosions in the face of the difference principle. For
in considering Hawking radiation, the whole system considered
necessarily includes the black hole \emph{together with} its
surroundings, i.e. space-time outside its event horizon.

It was supposed that the black hole of given mass might represent
a lower bound for a measure of the constructive evolvability, due
to internal differences, of a system of that mass. That may be the
case, if there is any way to consider a black hole as an isolated
system. But the picture of a 1kg black hole, with temperature of
the order $10^{31} K$ and rising, as a \emph{figure} against a
\emph{background} of relatively cold, empty space, is a picture of
extreme difference. Again, the difference principle actually leads
one correctly to expect that a great deal of constructive dynamics
are likely to ensue.

\section{Conclusion}
The difference principle of entropy dynamics has been presented as
a conceptual framework to provide insight into problems of
dynamics in general. As ``difference'' is a fundamentally
information-theoretic concept, this constitutes a viewpoint
founded in information accounting rather than in energy
accounting. Just as \emph{entropy}, as multiplicity of possible
states, and \emph{action}, as multiplicity of possible paths, can
be regarded as duals, so \emph{difference}, as a collective
measure of system non-equilibrium, and \emph{flow}, as a
collective measure of system change, may be regarded as
self-propagating duals. It has been shown that the difference
principle may provide some intuitive understanding and possible
insight even when pushed to its limit in the strange realm of
black-hole thermodynamics.

\begin{acknowledgments}
The author is grateful to Education New Zealand and the people of Aotearoa for their support of this work by the New Zealand International Doctoral Research Scholarship.
\end{acknowledgments}


\begin{thebibliography}{10}
\providecommand{\enquote}[1]{``#1''}
\expandafter\ifx\csname url\endcsname\relax
  \def\url#1{\texttt{#1}}\fi
\expandafter\ifx\csname urlprefix\endcsname\relax\def\urlprefix{URL }\fi

\bibitem{Solomonoff:1964}
R.~Solomonoff, \enquote{A formal theory of inductive inference,} \emph{Information and
  Control} \textbf{7}, 1--22 and 224--254 (1964).

\bibitem{Mermin:1998}
D.~Mermin, \enquote{What is quantum mechanics trying to tell us?,} \emph{American Journal
  of Physics} \textbf{66}, 753--767 (1998).

\bibitem{Wheeler:1990}
J.~Wheeler, \enquote{Information, physics, quantum: the search for links,} in
  \emph{Complexity, Entropy, and the Physics of Information}, edited by
  W.~Zurek, Perseus Books, 1990, vol. VIII of \emph{Proceedings, Santa Fe
  Institute Studies in the Sciences of Complexity}, pp. 3--28.

\bibitem{Feynman:1982}
R.~Feynman, \enquote{Simulating physics with computers,} \emph{International Journal of
  Theoretical Physics} \textbf{21}, 467--488 (1982).

\bibitem{Schrodinger:statistics}
E.~Schr\"{o}dinger, \emph{What is Life? The Physical Aspect of the Living
  Cell}, Cambridge University Press, Cambridge, 1944, pp. 8--17.

\bibitem{Anandan:2003}
J.~Anandan, \enquote{Laws, symmetries, and reality,} \emph{International Journal of
  Theoretical Physics} \textbf{42}, 1943--1955 (2003).

\bibitem{Nielsen:1987}
D.~Bennett, N.~Brene, and H.~Nielsen, \enquote{Random dynamics,} \emph{Physica Scripta}
  \textbf{T15}, 158--163 (1987).

\bibitem{Lanczos:1962}
C.~Lanczos, \emph{The Variational Principles of Mechanics}, University of
  Toronto Press, 1962, 2 edn.

\bibitem{Feynman:1948}
R.~Feynman, \enquote{Space--time approach to non--relativistic quantum mechanics,}
  \emph{Reviews of Modern Physics} \textbf{20}, 367--387 (1948).

\bibitem{Lotka:1922b}
A.~Lotka, \enquote{Natural selection as a physical principle,} \emph{Proceedings of the
  National Academy of Sciences} \textbf{8}, 151--154 (1922).

\bibitem{Cox:1946}
R.~Cox, \enquote{Probability, frequency, and reasonable expectation, }\emph{American
  Journal of Physics} \textbf{14}, 1--13 (1946).

\bibitem{Jaynes:1957}
E.~Jaynes, \enquote{Information theory and statistical mechanics,} \emph{Physical Review}
  \textbf{106}, 620--630 (1957).

\bibitem{Jaynes:1965}
E.~Jaynes, \enquote{Gibbs vs. Boltzmann entropies,} \emph{American Journal of Physics}
  \textbf{33}, 391--398 (1965).

\bibitem{Toffoli:2003}
T.~Toffoli, \enquote{A digital perspective and the quest for substrate-universal
  behaviors,} \emph{International Journal of Theoretical Physics} \textbf{42},
  147--151 (1982).

\bibitem{Caticha:2001}
A.~Caticha, \enquote{Entropic dynamics,} \emph{AIP Conference Proceedings} \textbf{617},
  302--313 (2001).

\bibitem{Caticha:2008}
A.~Caticha, \enquote{From inference to physics,} \emph{AIP Conference Proceedings}
  \textbf{1073}, 23--34 (2008).

\bibitem{Hofstadter:2002}
D.~Hofstadter, \enquote{Analogy as the core of cognition,} in \emph{The
  Analogical Mind}, edited by D.~Gentner, K.~Holyoak, and B.~Kokinov, MIT
  Press, Cambridge, MA, 2002, pp. 499--538.

\bibitem{Carnot:1977}
S.~Carnot, \enquote{Reflections on the Motive Force of Fire,} in
  \emph{Reflections on the Motive Force of Fire by Sadi Carnot and other Papers
  on the Second Law of Thermodynamics by E. Clapeyron and R. Clausius}, edited
  by E.~Mendoza, Peter Smith, Gloucester, MA, 1977, {E}nglish translation from
  original publication 1824.

\bibitem{Morowitz:1968}
H.~Mororwitz, \emph{Energy Flow in Biology; Biological Organization as a
  Problem in Thermal Physics}, Academic Press, 1968, pp. 1--15.

\bibitem{Spencer-Brown:motive}
G.~Spencer-Brown, \emph{Laws of Form}, George Allen and Unwin, London, 1969,
  p.~1.

\bibitem{Curie:1894}
P.~Curie, \enquote{On symmetry in physical phenomena, symmetry of an electrical
  field and of a magnetic field,} in \emph{Symmetries in Physics: Selected
  Reprints}, edited by J.~Rosen, American Association of Physics Teachers,
  1982, pp. 17--25, {E}nglish translation from original publication 1894.

\bibitem{Rosen:entropy}
J.~Rosen, \emph{Symmetry in Science}, Springer-Verlag, 1995, p. 145.

\bibitem{Onsager:1931}
L.~Onsager, \enquote{Reciprocal relations in irreversible processes,} \emph{Physical
  Review} \textbf{37}, 405--426 (1931).

\bibitem{Toffoli:1998}
T.~Toffoli, \enquote{How much of physics is just computation?,} \emph{Superlattices and
  Microstructures} \textbf{23}, 381--406 (1998).

\bibitem{Triantafyllou:1995}
M.~Triantafyllou, and G.~Triantafyllou, \enquote{An efficient swimming machine,}
  \emph{Scientific American} \textbf{272}, 64--70 (1995).

\bibitem{TriantafyllouEtAl:2004}
M.~Triantafyllou, A.~Techet, and F.~Hover, \enquote{Review of experimental work in
  biomimetic foils,} \emph{IEEE Journal of Oceanic Engineering} \textbf{29},
  585--594 (2004).

\bibitem{UsabEtAl:2004}
W.~Usab, J.~Hardin, and A.~Bilanin, \enquote{Bioinspired delayed stall propulsor,}
  \emph{IEEE Journal of Oceanic Engineering} \textbf{29}, 756--765 (2004).

\bibitem{BerdichevskyGitterman:1998}
V.~Berdichevsky, and M.~Gitterman, \enquote{Stochastic resonance and ratchets -- New
  manifestations,} \emph{Physica A} \textbf{249}, 88--95 (1998).

\bibitem{Penrose:thermodynamics}
R.~Penrose, \emph{The Road to Reality}, Alfred Knopf, 2005, pp. 686--734.

\bibitem{Bekenstein:1981}
J.~Bekenstein, \enquote{Universal upper bound on the entropy-to-energy ratio for bounded
  systems,} \emph{Physical Review D} \textbf{23}, 287--298 (1981).

\bibitem{Lloyd:2000}
S.~Lloyd, \enquote{Ultimate physical limits to computation,} \emph{Nature} \textbf{406},
  1047--1054 (2000).

\bibitem{Bekenstein:1973}
J.~Bekenstein, \enquote{Black holes and entropy,} \emph{Physical Review D} \textbf{7},
  2333--2346 (1973).

\bibitem{MargolusLevitin:1998}
N.~Margolus, and L.~Levitin, \enquote{The maximum speed of dynamical evolution,}
  \emph{Physica D} \textbf{120}, 188--195 (1998).

\bibitem{Hawking:1971}
S.~Hawking, \enquote{Gravitational radiation from colliding black holes,} \emph{Physical
  Review Letters} \textbf{26}, 1344--1346 (1971).

\bibitem{Hawking:1976}
S.~Hawking, \enquote{Balck holes and thermodynamics,} \emph{Physical Review D}
  \textbf{13}, 191--197 (1976).

\bibitem{Hawking:1974}
S.~Hawking, \enquote{Black hole explosions?}, \emph{Nature} \textbf{248}, 30--31 (1974).

\bibitem{LloydNg:2004}
S.~Lloyd, and Y.~Ng, \enquote{Black hole computers,} \emph{Scientific American}
  \textbf{291}, 52--61 (2004).

\end{thebibliography}
\end{document}